\documentclass{iaus}
\usepackage{graphicx}

\newcommand{\brg}{Br$\gamma$}
\newcommand{\micron}{$\mu$m}
\newcommand{\kms}{km~s$^{-1}$}

\title[Massive YSOs in high-mass star-forming regions ] %% give here short title %%
{Massive Young Stellar Objects in high-mass star-forming regions}

\author[Arjan Bik \etal]   %% give here short author list %%
{Arjan Bik $^1$, Lex Kaper $^2$, Wing-Fai Thi$^3$ \and Rens Waters$^2$ }

\affiliation{$^1$European Southern Observatory, Karl-Schwarzschild
           Strasse 2, Garching-bei-M\"unchen, D-85748, Germany \break
           email: abik@eso.org\\[\affilskip]
$^2$ University of Amsterdam, Kruislaan 403, 1098 SJ Amsterdam,
	   The Netherlands \break email:
           lexk@science.uva.nl, rensw@science.uva.nl\\[\affilskip]
$^3$ ESA research fellow, Research and Scientific Support Department, ESTEC, Keplerlaan
           1, 2201 AZ, Noordwijk, the Netherlands \break email:wthi@rssd.esa.int}

\pubyear{2005}
\volume{227}  %% insert here IAU Symposium No.
\pagerange{1--6}
\date{?? and in revised form ??}
\jname{Proceedings Massive Star Birth: A Crossroads of Astrophysics}
\editors{R. Cesaroni, E. Churchwell, M. Felli, \& C.M. Walmsley.}
\begin{document}

\maketitle

\begin{abstract}
 High-quality $K$-band spectra of point
sources, deeply embedded in massive star-forming regions, have
revealed a population of 20 young massive stars showing no
photospheric absorption lines, but only emission
lines. The $K$-band spectra exhibit one or more features commonly
associated with massive Young Stellar Objects surrounded by
circumstellar material: a very red color $(J-K) = 2$, CO bandhead
emission, hydrogen emission lines (sometimes doubly peaked), and FeII
and/or MgII emission lines. The CO emission comes from a relatively
dense ($\sim 10^{10} \mathrm{cm}^{-3}$) and hot ($T\sim 2000-5000$~K)
region, sufficiently shielded from the intense UV radiation field of
the young massive star. Modeling of the CO-first overtone emission
shows that the CO gas is located within 5 AU of the star. The hydrogen
emission is produced in an ionized medium exposed to UV radiation. The
best geometrical configuration is a dense and neutral circumstellar disk
causing the CO bandhead emission, and an ionized upper layer where the
hydrogen lines are produced. We argue that the
circumstellar disk is likely a remnant of the accretion via a
circumstellar disk.
\end{abstract}

\firstsection % if your document starts with a section,
\section{Introduction}

Although over the past decades significant progress has been made in
unraveling the formation process of low-mass stars (\cite[Shu \etal,
1987]{Shu87}), it is not well understood how massive stars form. The
contraction timescale is short, so that already very early in the
formation process massive stars will produce a copious radiation field
that may hamper or even reverse the accretion process
(e.g. \cite[Wolfire \& Cassinelli, 1987]{Wolfire87}). These
difficulties have led to the suggestion that stars more massive than
$\sim 10$~M$_{\odot}$ cannot form through (spherical) accretion alone,
but instead form by collisions of intermediate-mass stars
(\cite[Bonnell \etal, 1998]{Bonnell98}). Alternatively, non-spherical
accretion through a disk could solve the ``radiation pressure
problem'' (\cite[Yorke \& Sonnhalter, 2002]{Yorke02}). Therefore, the
detection and characterization of circumstellar disks around young massive stars is
regarded as an essential step in understanding the formation of the
most massive stars. Observational evidence is growing that rotating
circumstellar disks are present around high-mass protostars in the hot
core phase (\cite{Minnier98}, \cite{Shepherd01}, \cite{Beltran04},
\cite{Chini04}).

Already before arriving on the main sequence, the young massive star
produces a strong UV radiation field ionizing its surroundings. At
this stage the object will become detectable at far-infrared and radio
wavelengths through the heated dust and recombination of hydrogen in
the expanding hyper- or ultra-compact H~{\sc ii} region
(e.g. \cite[Churchwell, 2002]{Churchwell02}). In the mean time, the
same UV radiation field will impact on the extended circumstellar disk
and start to destroy it (\cite[Hollenbach et
al. 1994]{Hollenbach94}). The disk destruction timescale is so short
($\sim 100,000$~yr) that it becomes an observational challenge to
detect and measure any remnants of the formation process of massive
stars.

\begin{figure}
\includegraphics[width=1\columnwidth]{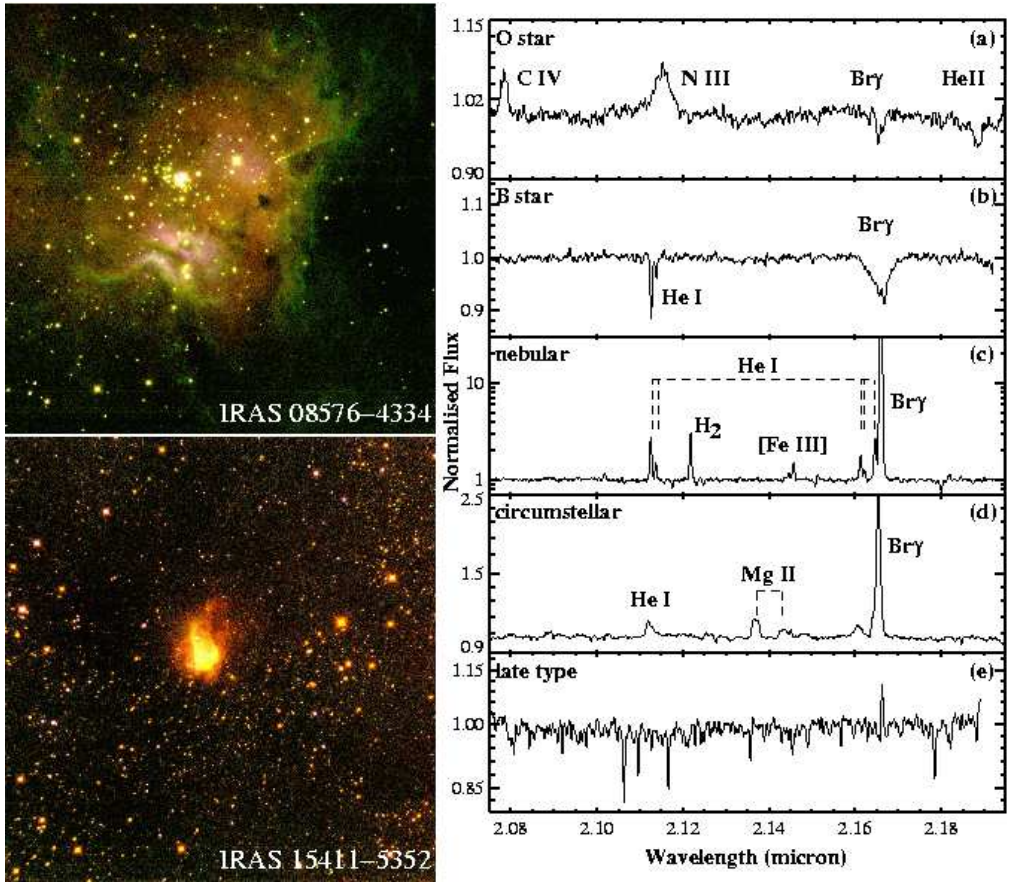}
 \caption{\emph{Left:} Near-infrared images of 2 of the high-mass
 star-forming regions in our sample. \emph{Right:} Examples of the
 different types of spectra obtained in our spectroscopic
 survey. Panels (a) and (b) show the photospheric $K$-band spectra of
 O and B stars.  Panel (c) shows a spectrum dominated by nebular
  emission from the UCHII region. The emission lines are narrow and
  not resolved. This in contrast to the spectrum displayed in panel
  (d) where the emission lines are resolved. This object
  shows characteristics of a massive YSO. Panel (e) displays a
  spectrum of a late type (fore-/background) star.}\label{fig:example}
\end{figure}

The near-infrared is the only wavelength range where the dust
obscuration is reduced and the dust emission is not dominant yet. This
favours the detection of the photospheric and circumstellar emission
of the young massive star. We have carried out an extensive
near-infrared survey of fields centered on southern ultra-compact
H~{\sc ii} regions detected at far-infrared and/or radio wavelengths,
with SOFI on the ESO {\it New Technology Telescope} (\cite[Bik,
2004]{Bikthesis}; \cite[Kaper \etal, 2005]{Kaper05}). These
narrow-band imaging observations have revealed embedded, young stellar
clusters ($A_{V} \sim 10-50$~magnitudes) containing newly formed
massive stars. (left panel of Fig. \ref{fig:example}).

% Follow-up K-band spectroscopy with ISAAC mounted on the ESO
%{\it Very Large Telescope} has confirmed their massive-star nature
%(\cite[Bik \etal, 2005a]{Bik05Ostar}). 

\subsection{The stellar content}

High-resolution (R=10,000) K-band spectroscopy of members of those
clusters have been obtained with ISAAC mounted on the VLT. The
spectroscopic targets have been selected based on their $K$-band
magnitude and $(J-K)$ color. Spectra have been taken of the
brightest and most reddened cluster members.  The obtained $K$-band spectra can
be divided in different classes (see right panel of
Fig. \ref{fig:example} for an example of the different spectra).
Forty objects turn out to be late-type (fore- or background) stars,
with $K$-band spectra dominated by many absorption lines
(Fig. \ref{fig:example}e). These stars are a natural by-product of our
selection criteria. They emit the bulk of their radiation in the
(near) infrared and have a red intrinsic color. The other three
classes are different types of objects connected with star-forming
regions. Thirty-eight objects show the spectral features indicative of
OB stars (Fig. \ref{fig:example}a,b). These stars are discussed in
\cite{Bik05a}.  The second type of objects are the near-infrared
counterparts of the UCHII regions with nebular $K$-band spectra
(Fig. \ref{fig:example}c).  We focus on the spectroscopic properties
of the third class; 20 objects that differ from normal main-sequence
OB stars and exhibit features commonly associated with massive Young
Stellar Objects (YSOs).

\section{Massive Young Stellar Objects}

Based on the K-band spectra of the massive YSOs candidates in our
sample we find evidence for the prescence of a circumstellar disk.
The massive YSO candidates show emission line spectra and are selected
based on the presence of \brg\ in their spectrum (\cite[Bik \etal,
2005b]{Bik05b}). These objects are the reddest members of their
respective clusters. This red color indicates that the object is
highly obscured and/or has a strong K-band excess. We cannot easily
separate these two contributions as we observed in the J and K-band
only. In some cases, however, we were able to determine the intrinsic
(J-K) of a neighboring star in the young cluster based on its K-band
spectrum, so that we can measure the interstellar extinction in that
sight-line. The \brg\ full width at half maximum (FWHM) is large (and
spectrally resolved), ranging from 100 to 230~\kms. The large FWHM
indicates that the emission has a circumstellar and not a nebular
origin. One object shows a double-peaked \brg\ profile, with a peak
separation of 95~\kms. Such a profile suggests formation in a rotating
disk. The \brg\ profile in several other targets displays an
asymmetry.

Some objects exhibit an Fe~{\sc ii} emission line (2.089~\micron),
and/or  show broad Mg~{\sc ii} emission (2.138,
2.144~\micron). The Mg~{\sc ii} emission is likely produced by the
excitation through Ly$\beta$ fluorescence. Like the Fe~{\sc ii}
emission, the Mg~{\sc ii} emission lines point to a dense and warm
(several 1000~K) circumstellar environment. In a few objects hydrogen
Pfund-lines are observed. The Pf-line emission indicates a line
forming region of ionized gas with high density ($N_\mathrm{e} \sim 10^{8}
\mathrm{cm}^{-3}$).

\subsection{CO bandheads}

\begin{figure}
\begin{center}
\includegraphics[width=0.9\columnwidth]{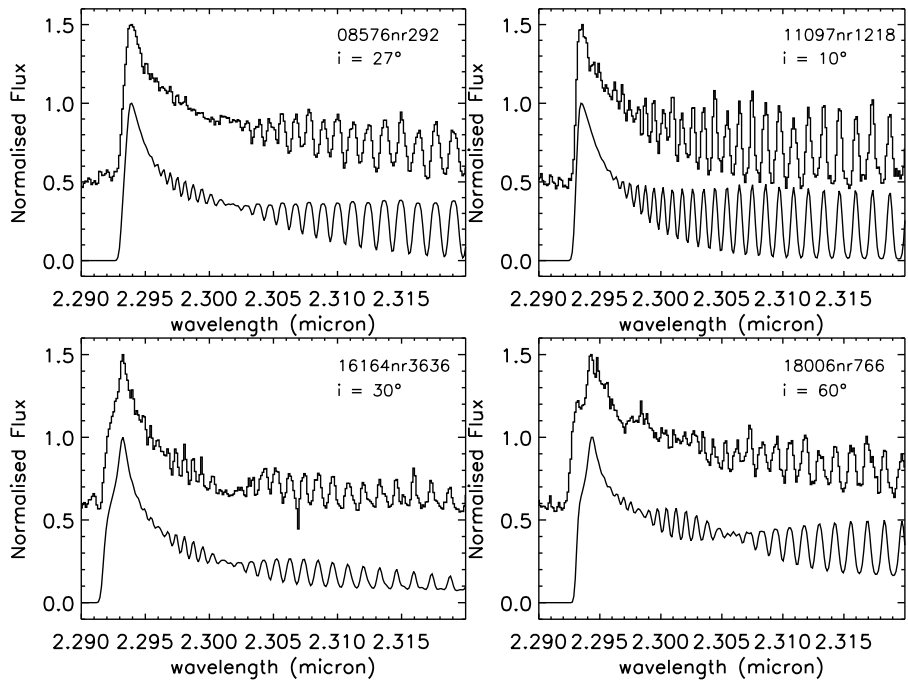}
\end{center}
 \caption{CO first-overtone bands in four objects (upper
spectra) shown along with the best-fitting model (lower spectra). The
object name and adopted inclination are given.}\label{fig2}
\end{figure}

%At the high densities and temperatures (1000-2000~K) characteristic of
%disks at 0.1--5~AU around young stars, molecules are expected to be
%sufficiently excited to produce a rich ro-vibrational spectrum in the
%near- (overtone transitions $\Delta v = \pm2$) and mid-infrared
%(fundamental transitions $\Delta v = \pm 1$). 

CO first-overtone
emission has been detected and analyzed in a number of high- and
low-mass stars (e.g. \cite[Scoville \etal, 1979]{Scoville79} and
\cite[Chandler \etal, 1995]{Chandler95}). Modeling
efforts have been hampered by the small number of observed $J$-lines,
or a low resolving power and moderate signal-to-noise.

Of our massive YSO sample, 15 objects are observed in the ``CO
setting'', of which 5 show the CO first-overtone bands in emission
(Fig.~\ref{fig2}).  The shape of the bandhead varies from source to
source.  The rise of the first bandhead is sharp in the spectrum of
08576nr292 and 11097nr1218 whereas in that of 16164nr3636 and
18006nr766 a blue wing is clearly present. The expected profile from a
spherical stellar wind has a flat-topped shape and cannot account for
the wing seen in two of the objects.  The favored model to explain the
blue wing consists of a keplerian disk and/or a disk-wind
(\cite[Chandler \etal, 1995]{Chandler95}), although the disk model
provides the best match.

In \cite{Bik04} we explore the possibility that the CO bandhead
emission arises from a disk in keplerian rotation around the massive
star. Synthetic CO bandhead spectra are generated using a standard
parametric disk model for a range of gas temperatures, column
densities, turbulent velocities and disk viewing angles. The
population of the CO rotational levels within each vibrational level
is assumed to be in local thermodynamic equilibrium, with an
excitation temperature given by the local vibrational temperature,
specific for each bandhead; a temperature gradient within the
line-forming region is neglected.  Typical column densities derived
from the model fitting are $10^{20}-10^{21}$~cm$^{-2}$ and excitation
temperatures are found to be in the range of 1700--4500~K.

The different observed profiles of the CO-bandhead emission can be
explained by a difference in inclination angle of the disks. It turns
out that the CO is located in the inner disk regions (within 5 AU),
very close to the central star (see \cite[Bik \& Thi, 2004]{Bik04} for
more details). The maximum distance at which the CO bandheads are emitted
depends on the inclination, which is not a well constrained parameter,
but even for a wide range of possible inclinations the hot CO
molecules are located within a few AU from the star. Similar results
are found by \cite{Blum04} who performed an analysis of a number of
massive YSOs located in giant HII regions.

In the absence of extinction by dust grains, the CO molecules should
be photodissociated by the stellar ultraviolet photons.  However, the
derived column densities (10$^{20}$ -- 10$^{21}$ cm$^{-2}$) are well
above the required value for the CO molecules to self-shield ($N$(CO)
$\sim 10^{15}$cm$^{-2}$, \cite[van Dishoeck \& Black,
1988]{vanDishoeck88}). Moreover, gas phase chemical models show that
CO molecules can rapidly form in the gas phase to compensate for their
destruction (\cite[Thi \& Bik, 2005]{Thi05}).

\subsection{Line-forming regions}

The different emission lines discussed above require different
physical conditions to be emitted. The CO first-overtone emission
requires the material to be neutral, dense and a temperature of
a few thousand K. On the other hand, the hydrogen emission requires
ionized material and to emit the Pfund lines a high electron density
is required ($N_\mathrm{e} \sim 10^{8} \mathrm{cm}^{-3}$). This means
that the different lines are formed at different locations in the
circumstellar environment.

The Pfund lines are much broader than \brg.  If the Pf-lines would
originate in a keplerian rotating disk, they are likely formed in the
high-density inner disk region, while \brg\ would be formed over a
more extended region of the disk. This would explain the difference
between the width Pf-lines and \brg. \brg\ will be dominated by
the more slowly rotating outer parts of the disk (larger surface area)
and will have a smaller FWHM.

Another possibility is that \brg\ is formed in a disk wind. The FWHM
of the \brg\ lines does not correlate with the width of the CO lines
of which modeling shows that these lines are formed in a circumstellar
disk. The FWHM of the \brg\ lines for the objects with the steep CO
bandheads is similar to those of the objects where the CO bandheads
are broad. In the disk wind scenario the outflow velocity is higher
than the rotation velocity and widths of $\sim 200$ \kms are expected
for lines formed in a disk wind (\cite[Drew \etal, 1998]{Drew98}); no
correlation between the FWHM of \brg\ and disk inclination is
expected.

\section{The nature of the cirumstellar mater}

\begin{figure}
\includegraphics[width=1\columnwidth]{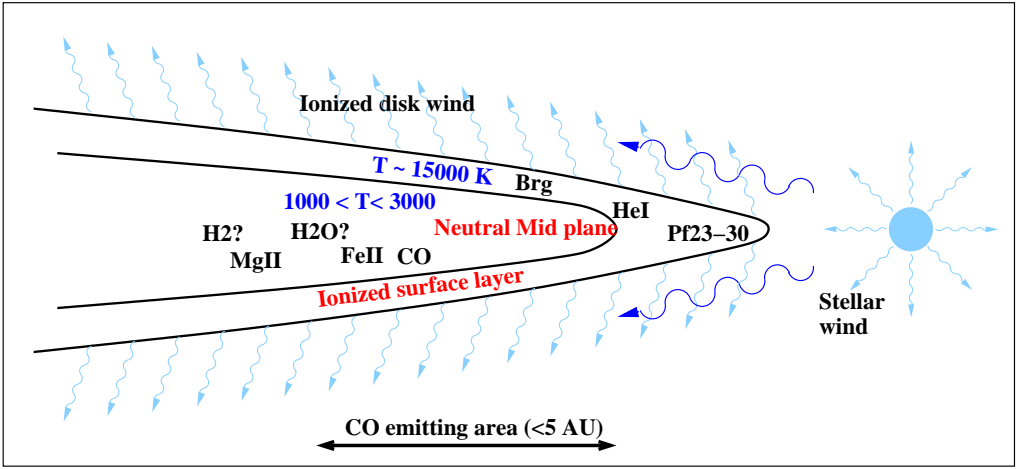}
 \caption{A schematic view of the different line-forming regions in
the circumstellar disk surrounding a massive YSO. The CO is located in
the dense mid plane of the disk, while the hydrogen lines are emitted
by the ionized surface layer and/or disk wind.}\label{fig3}
\end{figure}

We have presented several arguments that the emission-line objects
contained in our sample are surrounded by a dense, circumstellar disk
in keplerian rotation: broad and sometimes double-peaked hydrogen
emission lines, Mg~{\sc ii} emission, CO first-overtone band emission
likely arising from a keplerian disk and a strong near-infrared
excess. Fig.~\ref{fig3} gives a schematic view of the disk geometry
and the proposed locations of the different line forming regions.

After correction of the interstellar extinction, the location of the
massive YSO in the (J-K) vs K color-magnitude diagram suggests that
they are of early spectraltype. The majority of the objects have
photometric properties similar to the Herbig Be stars, suggesting that
those objects are likely B-type stars.  \cite{Hanson97} were able to
detect the I-band spectrum of some of the massive YSOs in M17
resulting in a B-star classification. They found that the O stars in
M17 do not show evidence for circumstellar disks. This is confirmed by
observations of massive stars in giant H~{\sc ii} regions by
\cite[Blum \etal\ (1999]{Blum99}, \cite[2000]{Blum00},
\cite[2001)]{Blum01} and \cite[Figuer{\^e}do et al
(2002]{Figueredo02}, \cite[2005)]{Figueredo05}. In our sample only a
few objects show evidence that the central star is of O spectral type.

The disks around the objects in our sample might be the remnants of
the large accretion disks ($\sim 10\,000$~AU) detected in the
millimeter around younger high-mass protostars.  Although the near-infrared
only probes material upto a few AU of the star, the disks around the
objects in our sample are likely much smaller than those around the
high-mass proto stars ($\sim$ 1000 AU).  The FUV radiation and the
strong stellar wind of the stars start to rapidly disrupt the inner
regions and photo-evaporate the outer regions of the circumstellar
disk as soon as it has formed (e.g. \cite[Hollenbach \etal,
2000]{Hollenbach00}).  The detection of only a few O stars surrounded
by circumstellar matter is consistent with these photo-evaporation and
disk destruction models as well.  The disks around O stars are
destroyed much more rapidly than those around B stars. This suggests
that the few objects for which the central star is likely an O star are
the youngest objects in our sample (few 100,000 years).

\end{document}